# Artificial-Intelligence-Based Triple Phase Shift Modulation for Dual Active Bridge Converter with Minimized Current Stress

Xinze Li, *Student Member*, *IEEE*, Xin Zhang, *Senior Member*, *IEEE*, Fanfan Lin, *Student Member*, *IEEE*, Changjiang Sun, *Member*, *IEEE*, Kezhi Mao, *Member*, *IEEE*

*Abstract*—The dual active bridge (DAB) converter has been popular in many applications for its outstanding power density and bidirectional power transfer capacity. Up to now, triple phase shift (TPS) can be considered as one of the most advanced modulation techniques for DAB converter. It can widen zero voltage switching range and improve power efficiency significantly. Currently, current stress of the DAB converter has been an important performance indicator when TPS modulation is applied for smaller size and higher efficiency. However, to minimize the current stress when the DAB converter is under TPS modulation, two difficulties exist in analysis process and realization process, respectively. Firstly, three degrees of modulation variables in TPS modulation bring challenges to the analysis of current stress in different operating modes. This analysis and deduction process leads to heavy computational burden and also suffers from low accuracy. Secondly, to realize TPS modulation, if a lookup table is adopted after the optimization of modulation variables, modulation performance will be unsatisfactory because of the discrete nature of lookup table. Therefore, an AI-based TPS modulation (AI-TPSM) strategy is proposed in this paper. Neural network (NN) and fuzzy inference system (FIS) are utilized to deal with the two difficulties mentioned above. With the proposed AI-TPSM, the optimization of TPS modulation for minimized current stress will enjoy high degree of automation which can relieve engineers' working burden and improve accuracy. In the end of this paper, the effectiveness of the proposed AI-TPSM has been experimentally verified with a 1 kW prototype.

## I. INTRODUCTION

The dual active bridge (DAB) isolated bidirectional DC-DC converter has been widely adopted in many applications since it was firstly proposed in 1992 [1], such as wireless power transfer, electric vehicles, DC microgrid [2] and solid state transformer [3]. This topology consists of one high-frequency transformer and two full bridges, as shown in Fig. 1. It attracts much attention for its galvanic isolation, high power density and bidirectional power transfer capability [4], [5].

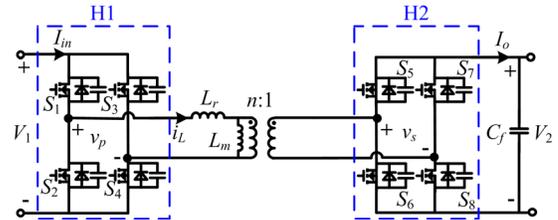

Fig. 1. Typical schematic of an isolated DAB converter with single inductor $L$.

In terms of the modulation strategies for DAB converter, phase shift modulation is most commonly used because of its simple implementation and fundamental frequency operation [6]. The simplest technique in phase shift modulation is single phase shift (SPS) modulation, which has one degree of control freedom in the phase shift between two full bridges [7]. By modifying SPS modulation, extended phase shift (EPS) adds one more degree of control freedom in the duty cycle of one full bridge [8], [9]. Similarly, on the basis of SPS, dual phase shift (DPS) modulation makes the duty cycle of two full bridges controllable and share the same value [10], [11]. To further improve modulation performance, triple phase shift (TPS) modulation has been introduced to include one more degree of control freedom, which realizes three degrees of control freedom in total [12]–[15]. It can control the duty cycle of two full bridges and the phase shift between them, and thus SPS, EPS, and DPS can be regarded as special cases of TPS. With all possible modulation variables, TPS modulation widens zero voltage switching (ZVS) range and improves power efficiency [16]. Hence, TPS can be viewed as one of the most advanced techniques in phase shift modulation.

Currently, the optimization of current stress is a popular trend in the TPS modulation for DAB converter. In this paper, peak inductor current is regarded as the current stress objective to be optimized, which brings advantages to many aspects. First of all, switching losses are bounded up by peak current, so the optimization of which can reduce switching losses [17]. In addition, the optimization of peak current can reduce root-mean-square current and improve efficiency [18]. Moreover, optimal peak current stress can reduce the size of magnetic core and protect power devices [19], [20].

However, due to the complexity brought by three degrees of control freedom in TPS modulation, the analytical formula of current stress is considerably difficult to derive [21], [22]. In previous research works, researchers have investigated the

Manuscript received March 27, 2021; revised May 12, 2021; revised July 05, 2021; accepted August 04, 2021. This work was supported by Start-up grant of Professor Zhang at Zhejiang University. (*Corresponding author: Xin Zhang*).

Xinze Li and Kezhi Mao are with the School of Electrical and Electronic Engineering, Nanyang Technological University, Singapore 639798, Singapore. (e-mail: xinze001@e.ntu.edu.sg; EKZMao@ntu.edu.sg).
Xin Zhang is with the College of Electrical Engineering, Zhejiang University, Hangzhou 310027, China, and with Hangzhou Global Scientific and Technological Innovation Center, Zhejiang University, Hangzhou 310058, China. (e-mail: zhangxin_ieee@163.com).
Fanfan Lin is with ERI@N, Interdisciplinary Graduate Program, Nanyang Technological University, Singapore 639798, Singapore. (e-mail: fanfan001@e.ntu.edu.sg).
Changjiang Sun is with Rolls-Royce@NTU Corporate Lab, Nanyang Technological University, Singapore 639798, Singapore. (e-mail: changjiang.sun@ntu.edu.sg).

waveforms in every operating mode and integrated the current piece by piece, which is time-consuming [6], [12], [22]. With the derived analytical formula, which is very complex, optimization with these manually-deduced mathematical expressions also face challenges. Besides high computational complexity, some assumptions are supposed before the derivation process, such as the assumption of lossless components and negligible magnetizing current [18], [23]. And these assumptions and approximations undermine the accuracy of the analytical results, even though they simplify the analysis.

In the aspect of realizing TPS modulation strategy, there are usually two ways. One method is to store the derived formula of optimal results in the modulator [18], [22], [23]. Modulation results will be calculated according to practical conditions. As discussed above, the method based on analytical formula is easy to implement, whereas it suffers from complicated deduction process and time-consuming problem. The other method is to store the optimal modulation results in a lookup table instead of deriving analytical formula [24]–[26]. In practice, when current operating conditions have been detected, the modulation variables in this lookup table will be searched and then applied. However, the modulation results provided by the lookup table are discrete, which leads to a case when the practical specifications are not listed in this table.

To overcome the difficulties in the optimization of TPS modulation for DAB converter, researchers have considered to apply some AI tools. Tang et.al. in [26] have tried to mitigate human-dependance in the optimization process. They have utilized Q-learning to find optimal modulation variables, which are stored in a lookup table. However, analytical process to derive formula with approximations remains and the lookup table brings discrete modulation results. Moreover, Harrye et.al have adopted neural network (NN) as a TPS controller to minimize reactive power [27]. But the reactive power still needs to be deduced manually and the modulation performance is not smooth as expected because the NN controller is applied in an open loop control.

Aimed to optimize current stress of DAB converter under TPS modulation, an AI-based TPS modulation (AI-TPSM) strategy is proposed in this paper. Generally, AI-TPSM contains three stages with three different AI tools. Firstly, NN is trained with simulations to learn the relationships between current stress and variables, which is to replace traditional complicated and inaccurate analytical process. Secondly, particle swarm optimization (PSO) algorithm is adopted to find the optimal modulation results which can minimize current stress. Lastly, fuzzy inference system (FIS) is utilized to store the optimal modulation results under different operating conditions, which can provide continuous modulation. The proposed AI-TPSM enjoys high degree of automation which can relieve engineers' working burden and improve accuracy.

This paper is structured as follows: after the introduction, operation principles of TPS modulation and the existing challenges will be described in Section II; the detailed process of AI-TPSM is illustrated in Section III; an application case of AI-TPSM is given in Section IV; hardware experimental results are displayed in Section V and conclusion is summarized in Section VI.

## II. OPERATING PRINCIPLE OF TPS MODULATION AND THE EXISTING CHALLENGES

### A. Operating Principle of TPS modulation for DAB Converter

The circuit configuration of DAB converter is presented in Fig. 1. Two full bridges $H_1$ and $H_2$ are connected with a magnetic tank which includes an inductor $L$ and a high frequency transformer. The ac voltages generated by $H_1$ and $H_2$ are $v_p$ and $v_s$, respectively. Under TPS modulation, gate driving signals and waveforms of $v_p$ and $v_s$ are depicted in Fig. 2.

In Fig. 2, $2T$ is a complete switching period $T_s$. $D_0$ is the phase shift between two full bridges ($S_1$ and $S_5$) which belongs to [-1,1]. $D_1$ and $D_2$ are the duty cycle of $v_p$ and $v_s$, respectively. The ranges of $D_1$ and $D_2$ are both in [0,1]. All of the three degrees of control freedom can be adjusted to control the transferred power and the inductor current $i_L(t)$. According to [22], the transferred maximum power can be expressed as (1), where $f_s$ is the switching frequency. Based on the maximum power to be transmitted $P_{max}$, the value of single inductor $L$ follows (2).

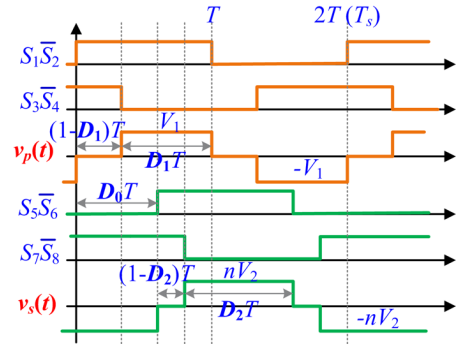

Fig. 2. Operating principles of TPS modulation.

$$P_{\max} = \frac{nV_1V_2}{8f_sL} \quad (1)$$

$$L \leq \frac{nV_1V_2}{8f_sP_{\max}} \quad (2)$$

### B. Challenge Descriptions for Optimization of TPS Modulation with Minimized Current Stress

The process to achieve optimal TPS modulation with minimized current stress and carry out this modulation in practice can be generally divided into three stages, as shown in Fig. 3.

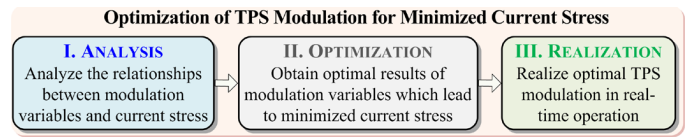

Fig. 3. The process to achieve optimal TPS modulation with minimized current stress and carry out the modulation.

After the specification of operating conditions, the relationships between current stress and variables are analyzed. Variables include modulation variables ($D_0$, $D_1$ and $D_2$) and operating conditions (output power $P$ and output voltage $V_2$), and current stress is represented by the peak $i_{pk}$ of the current through inductor $i_L$. And then the modulation variables will be optimized for minimal current stress under different operating conditions. In this last stage, the optimized modulation variables will be implemented in real time according to different operating conditions.

However, in this process, some challenges exist in Stage I and Stage III which increase complexity and undermine accuracy, as shown in Fig. 4.

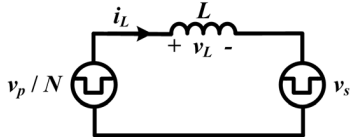

Fig. 4. Challenge descriptions for optimization of TPS modulation.

*(a) Challenge in Stage I: Analysis of Current Stress*

Fig. 5. Equivalent circuit of DAB converter with a single $L$.

The analysis of current stress in previous research papers follows the process below [18], [19], [22], [23]. The equivalent circuit in Fig. 5 is utilized to analyze the inductor current $i_L(t)$ and its peak value $i_{pk}$ (current stress) under all the switching modes of TPS modulation. For given values of $D_0$, $D_1$ and $D_2$, by applying the principle of inductor volt-second balance piecewise, $i_L(t)$ and $i_{pk}$ can be analyzed segment by segment within a switching period [22], [23].

This process suffers from two drawbacks. Firstly, the manual derivation of the expressions of $i_L(t)$ and $i_{pk}$ under all the switching modes is tedious and complex. This process is described with Fig. 6. TPS modulation has totally three modulation variables and 12 switching modes considering both power transfer directions, and thus the derived expression current stress $i_{pk}$ has high degree of complexity, which has been presented in [18], [22], [28]. Secondly, accuracy is undermined during analysis process because of the assumptions of the lossless component and negligible magnetizing current.

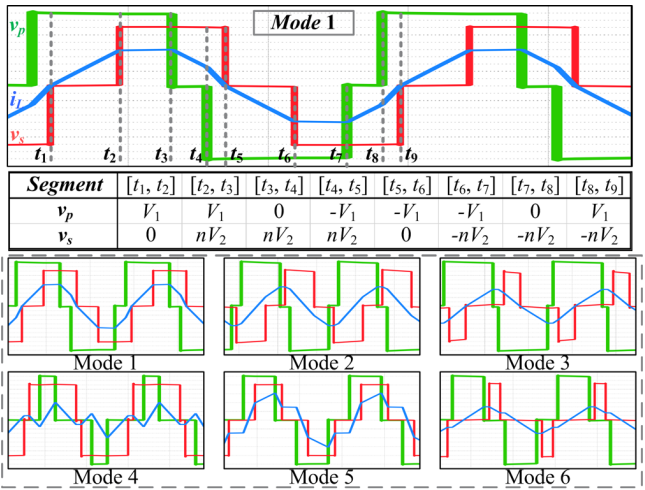

Fig. 6. Challenge in Stage I: complex segment-by-segment analysis of current stress under different operating modes of TPS modulation.

*(b) Challenge in Stage III: Realization of TPS Modulation*

To realize TPS modulation in practical applications, the optimal values of modulation variables which are obtained in Stage II will be stored in a lookup table. This lookup table will be saved in the memory of microcontroller. In real-time operation, once the current operation conditions ($P$ and $V_2$) are detected, corresponding optimal values of modulation variables ($D_0$, $D_1$ and $D_2$) will be called from the lookup table [23].

Due to the nature of lookup table as shown in Fig. 7, the data saved is discrete. As a result, a situation when ($P$, $V_2$) lies in between ($P$, $V_2$)$_i$ and ($P$, $V_2$)$_{i+1}$ may occur. In this situation, either ($P$, $V_2$)$_i$ or ($P$, $V_2$)$_{i+1}$ will be considered as approximation, whose ($D_0$, $D_1$, $D_2$) are actually not the optimal one for that specific operating condition [26]. Chances are that the obtained modulation variables fail to present optimal performance, deteriorating the accuracy.

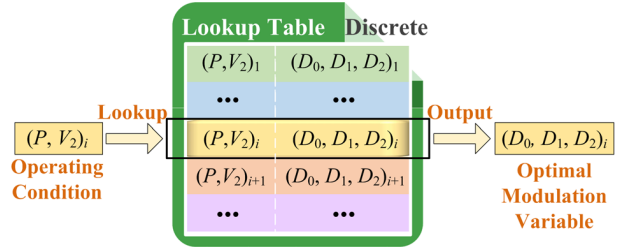

Fig. 7. Challenge II: discrete nature of lookup table for the realization of TPS modulation.

In summary, the existing optimization approaches of TPS modulation for minimal current stress have some challenges in high complexity and low accuracy which need to be addressed.

## III. THE PROPOSED AI-BASED TPS MODULATION

To solve the challenges described above, an AI-based TPS modulation (AI-TPSM) optimization strategy to minimize current stress is proposed in this paper. Generally, AI-TPSM contains three stages with one specific AI tool in every stage. The entire process of AI-TPMS optimization is described with Fig. 8.

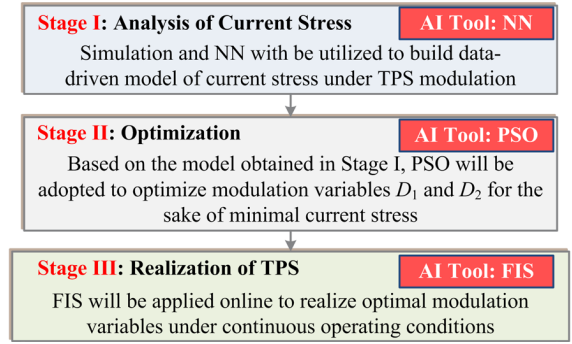

Fig. 8. Descriptions for the proposed AI-TPSM.

### A. Stage I: NN-Based Analysis of Current Stress

Aimed at the challenge of complicated manual analysis, data-driven model of current stress is achieved via NN in this paper, realizing high-level automation in analysis. The detailed process of Stage I is given in Fig. 9 and illustrated as the following.

Before the start of Stage I, all the operating specifications should be firstly decided, including the input voltage $V_1$, output voltage $V_2$, output power $P$ and switching frequency $f_s$.

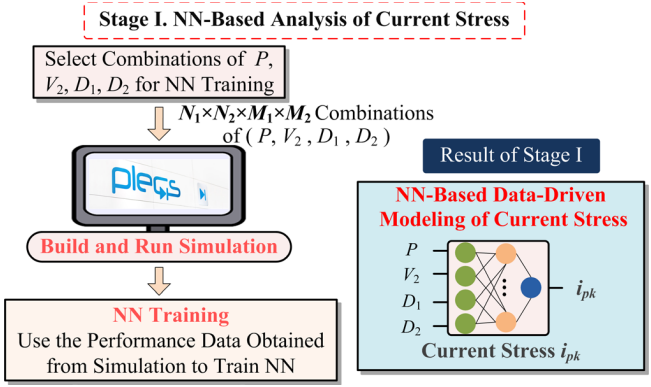

Fig. 9. Flowchart of Stage I: analysis of current stress with NN.

Firstly, combinations of operating conditions ($P$, $V_2$) and modulation variables ($D_1$, $D_2$) are specified. For operating conditions $P$ and $V_2$, $N_1$ number of values of $P$ and $N_2$ number of values of $V_2$ are evenly selected within [$P_{min}$, $P_{max}$] and [$V_{2\_min}$, $V_{2\_max}$], respectively. As introduced in Section II, modulation variables $D_1$ and $D_2$ are both in [0, 1]. $M_1$ number of values of $D_1$ and $M_2$ number of values of $D_2$ are evenly selected within [0, 1]. As a result, the total number of combinations of operating conditions and modulation variables is $N_1 \times N_2 \times M_1 \times M_2$.

After that, simulation model of the DAB converter under TPS modulation is built in PLECS software. This simulation will be implemented for the total $N_1 \times N_2 \times M_1 \times M_2$ combinations of operating conditions ($P$, $V_2$) and modulation variables ($D_1$, $D_2$) to collect performance data for current stress. In the simulation, to regulate power transfer flow and to maintain stable output voltage of DAB converter under TPS modulation, $D_0$ is automatically determined by the PI regulator for the given combination of $P$, $V_2$, $D_1$ and $D_2$ [18].

Afterwards, based on the current stress performance data collected with simulation model, NN can be trained. The total $N_1 \times N_2 \times M_1 \times M_2$ number of simulation data is divided into training set (70%), validating set (15%) and testing set (15%), which are used for training NN, selecting NN structure and testing the trained NN on new and unseen data points. After the training of NN, it can serve as a data-driven model of current stress which can evaluate current stress performance for any possible combination of operating conditions and modulation variables. It should be noted that current stress is evaluated by the peak current through inductor $L$ ($i_{pk}$).

### B. Stage II: Optimization with PSO Algorithm

In Stage II, based on the data-driven model built in Stage I, optimal modulation variables $D_1$, $D_2$ under different operating conditions $P$, $V_2$ will be found.

The mathematical formulation of the optimization problem can be expressed as follows:

**For the given operating conditions $P$ and $V_2$, the goal is:**

$$i_{pk}^* = \min_{D_1, D_2} i_{pk}(P, V_2, D_1, D_2) \quad (3)$$

**Subject to:**

$$0 \leq D_1 \leq 1 \quad (4)$$
$$0 \leq D_2 \leq 1 \quad (5)$$

To satisfy the power transfer and voltage regulation requirements of TPS modulation, $D_0$, as one of the modulation variables, will be determined by the output of PI regulator once $D_1$ and $D_2$ are specified. Hence, $D_0$ is not an independent optimization variable to be considered.

To solve this optimization problem, particle swarm optimization (PSO) algorithm is chosen to get the optimal modulation variables $D_1$ and $D_2$. PSO algorithm is an evolutionary algorithm which mimics the behavior of bird flocks to search for optimal results in the solution space [29], [30]. To solve the optimization problem in (3), dimension of the particle position is set as 2, consisting of two modulation variables ($D_1$, $D_2$). The position $X$ of the particle represents the values of $D_1$ and $D_2$. The velocity $V$ of the particle is the change of position in every iteration. The velocity of each particle in the 2-dimension space is updated with (6) and the position is updated with (7):

$$V_i^{m+1} = \omega V_i^m + c_1 r_1 (P_{best i}^m - X_i^m) + c_2 r_2 (G_{best}^m - X_i^m) \quad (6)$$

$$X_i^{m+1} = X_i^m + V_i^m \quad (7)$$

where $m$ is the iteration number, $c_1$ and $c_2$ are the learning coefficients, $\omega$ is the inertia weight, $P_{best}$ is the personal best information and $G_{best}$ is the global best information.

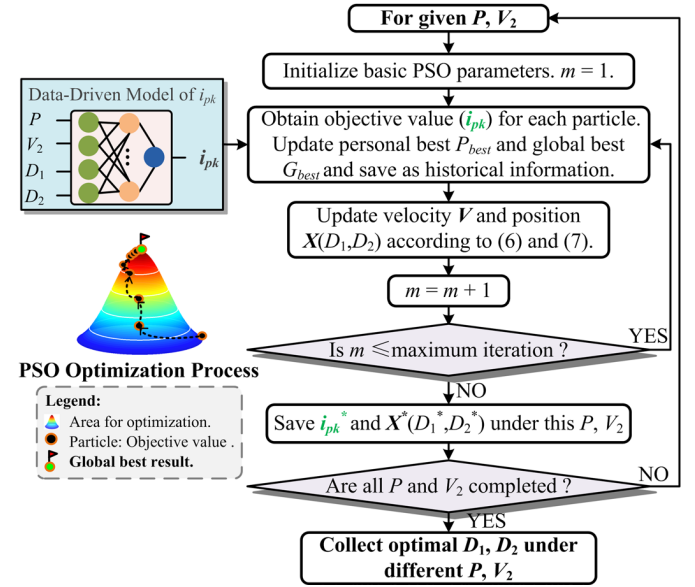

Fig. 10. Flowchart of Stage II: PSO for optimization.

The entire process of Stage II to find optimal modulation variables $D_1$ and $D_2$ under different operating conditions is described with Fig. 10. Firstly, operating conditions $P$ and $V_2$ are given. After the initialization, objective value $i_{pk}$ is evaluated for all particles individually in every iteration. And the position and velocity of every particle will be updated according to (6) and (7). Then, the found optimal $i_{pk}^*$ and the corresponding modulation variables $D_1^*$ and $D_2^*$ under the given $P$, $V_2$ are saved when the stopping criterion has been met. This process repeats until the optimal $D_1$ and $D_2$ under all combinations of $P$ and $V_2$ have been found.

With the PSO algorithm in Stage II, optimal $D_1$ and $D_2$ can be obtained for the given combinations of $P$ and $V_2$ to realize minimized current stress performance.

## C. Stage III: *Realization of TPS with FIS*

To avoid the inaccurate optimal modulation variables caused by discrete lookup tables, fuzzy inference system (FIS) is adopted, achieving continuous and accurate TPS modulation in real-time applications. FIS is chosen rather than standard interpolation techniques such as cubic interpolation because of its good interpretability and superior generalization capability [31]. The FIS-based control diagram for DAB under TPS modulation as proposed in this paper is given in Fig. 11.

To meet the requirements of power transfer and voltage regulation under TPS modulation, one of the three modulation variables ($D_0$, $D_1$, $D_2$) should be tuned by PI regulator. In the proposed FIS-based closed-loop control diagram for DAB under TPS modulation, the phase shift $D_0$ between two full bridges is chosen as the output of PI module, the input of which is the error between the reference voltage $V_{2,ref}$ and $V_2$.

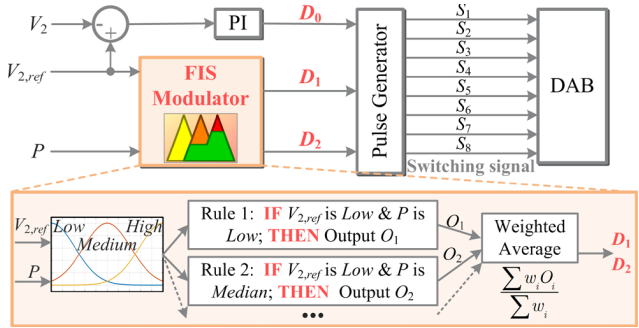

Fig. 11. Stage III: online realization of TPS with FIS.

As for the duty cycles $D_1$ and $D_2$ inside two full bridges, FIS modulator decides those values. The principle of FIS to realize online TPS modulation is shown in Fig. 11. In FIS modulator, input membership functions will compute the degree of membership that $V_{2,ref}$ and $P$ belong to each of the linguistic set (e.g., $P$ is high, $V_{2,ref}$ is low). Afterwards, all fuzzy rules will be applied to compute their outputs $O_1, \ldots O_N$, where $N$ is the total number of rules. Finally, $D_1$ and $D_2$ will be computed by the weighted average of all the outputs, in which the weights are decided by the firing strength of each rule [32].

|  | Continuity | Time Complexity | | Space Complexity | |
|---|---|---|---|---|---|
| LUT | Discrete | Low | Dependent on Data Size | Large | Dependent on Data Size |
| FIS | Continuous | Low | Independent of Data Size | Low | Independent of Data Size |

Fig. 12. Comparisons between LUT and the adopted FIS.

With the FIS-based modulation in Stage III, modulation variables $D_1$ and $D_2$ can be adaptively tuned in online applications to realize minimal current stress under varying operating conditions $P$ and $V_2$. As shown in Fig. 12, compared with the discrete lookup tables (LUT), FIS-based TPS modulation can realize optimal modulation variables $D_1$ and $D_2$ under the practical situations of varying operating conditions $P$ and $V_2$ with continuous values. In terms of time complexity, both LUT and FIS have fast computation speed. As for space complexity, LUT requires large storage space, while FIS has low space complexity. Moreover, the complexity of FIS is independent of data size, while the complexity of LUT rises with the increasing of data size [33].

In the proposed AI-TPSM, Stage I trains NN-based data-driven model of current stress, mitigating the dependence of tedious manual current stress deduction. PSO is utilized in Stage II to minimize the current stress for given combinations of operating conditions. FIS-based control diagram of TPS modulation is proposed in Stage III to obtain optimal modulation variables under continuous operating conditions in real-time applications.

## IV. DESIGN CASE OF APPLYING THE PROPOSED AI-TPSM

With the proposed AI-TPSM approach elaborated in Section III, a current-stress-optimal TPS modulation strategy for DAB converter is designed. The design case is illustrated as the following stage by stage.

In this design case, operating conditions and requirements of the DAB converter under TPS modulation are listed in Table I, where the output power $P$ can vary from 100W to 1000W, and output voltage $V_2$ can vary from 160V to 230V. Out of safety reasons, C2M0080120D is chosen. Even if other power switches are considered, the proposed AI-TPSM is still applicable.

In this paper, for the sake of illustration convenience and computation feasibility, only $P$ and $V_2$ are considered as varying operating conditions. If input voltage $V_1$ is also considered, AI-TPSM can still be applied by incorporating the varying of $V_1$ into simulation and NN training (Stage I), optimization with PSO (Stage II) and FIS-based online realization (Stage III).

TABLE I. DESIGN SPECIFICATIONS

| Rated Operating Specifications | | | |
|---|---|---|---|
| $P$ | 1000 W | $V_2$ | 200 V |
| $V_1$ | 200 V | $f_s$ | 20 $kHz$ |
| **Power Switches** | | | |
| Switches | C2M0080120D, Cree | Dead time | 500 $n$s |
| $R_{DS(on)}$ | 80 $m\Omega$ | $V_{DSS}$ | 1200 V |
| **High-Frequency Transformer** | | | |
| Inductor $L$ | | 140 $\mu$H | |
| Core material | | Iron based nanocrystalline alloy | |
| **Modulation Variables** | | | |
| Duty cycle of full bridge 1 $D_1$ | | Duty cycle of full bridge 2 $D_2$ | |
| **Operating Conditions** | | | |
| Output power $P$ | | Output voltage $V_2$ | |
| **Ranges of Modulation Variables** | | | |
| $D_1$ | | $D_{1\ min} = 0$; $D_{1\ max} = 1$ | |
| $D_2$ | | $D_{2\ min} = 0$; $D_{2\ max} = 1$ | |
| **Ranges of Operating Conditions** | | | |
| $P$ | | $P_{min} = 100$ W; $P_{max} = 1000$ W | |
| $V_2$ | | $V_{2\ min} = 160$ V; $V_{2\ max} = 230$ V | |

### A. Stage I: *NN-Based Analysis of Current Stress*

In Stage I of the proposed AI-TPSM approach, by following the flowchart in Fig. 9, NN-based data-driven model of current stress is automatically deduced. The steps are summarized as follows.

- Firstly, 20×20×20×20 (160000 in total) combinations of $P$, $V_2$, $D_1$ and $D_2$ are evenly selected to properly cover their ranges for the benefits of NN training.
- Subsequently, the simulation of DAB converter under TPS modulation is built using PLECS software and is repeated for all the given combinations of $P$, $V_2$, $D_1$ and $D_2$. Current stress performance is obtained in each simulation.

- All the simulation-generated performance data is partitioned into training set (70%), validating set (15%) and testing set (15%). NN-based data-driven model of current stress can be trained on training set, where the inputs are $P$, $V_2$, $D_1$ and $D_2$, and output is the current stress indicator $i_{pk}$. The structure of neural network is chosen according to the lowest error on validating set, as given in Table II and Fig. 13. Compared with other regression techniques [34] such as response surface, Bayesian regression and support vector regression (SVR), the trained NN manifests superior accuracy on all the datasets. As shown in Table III, the average percentage deviations of the trained NN are much smaller than other techniques. Even in the worst-case scenarios, NN still gives satisfactory accuracy in all the considered datasets.

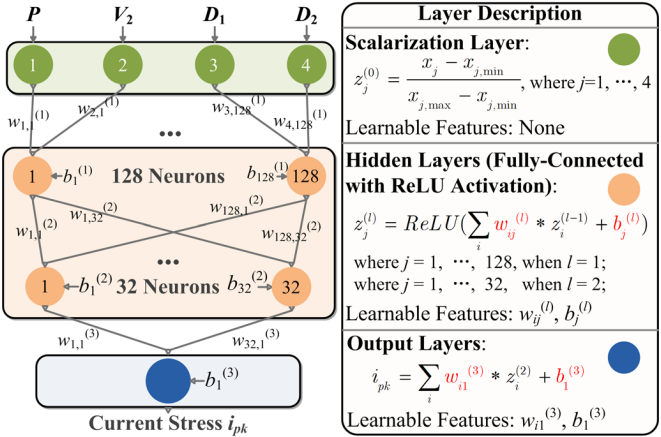

Fig. 13. Structure of the selected NN.

TABLE II. CONFIGURATION OF NN AND ITS OPTIMIZER

| Selected NN | |
| --- | --- |
| Inputs | $P$, $V_2$, $D_1$, $D_2$ |
| Output | Current stress $i_{pk}$ |
| Hidden Layers | Two layers, each of which have 128 and 32 neurons with ReLU activations |
| **NN Optimizer** | |
| Data for NN Training | 70%, 112,000 |
| Data for Structure Selection | 15%, 24,000 |
| Data for Testing New Data | 15%, 24,000 |
| Optimizer | Adaptive subgradient method [35] |
| Learning Rate | 0.001 |
| Regularization Coefficient | 1e-5 |
| Maximal Epochs | 10,000 |

TABLE III. ACCURACY OF THE TRAINED NN

| Percentage Deviations | Training Set | | Validating Set | | Testing Set | |
| --- | --- | --- | --- | --- | --- | --- |
| | Average | Largest | Average | Largest | Average | Largest |
| Response Surface | 26.2% | 42.0% | 26.5% | 42.6% | 26.5% | 39.3% |
| Bayesian Regression | 25.6% | 38.6% | 25.8% | 40.2% | 25.7% | 41.4% |
| SVR | 9.14% | 13.1% | 9.07% | 12.3% | 9.1% | 12.9% |
| **NN** | **0.47%** | **1.39%** | **0.46%** | **1.33%** | **0.45%** | **1.29%** |

### B. Stage II: Optimization with PSO Algorithm

In Stage II, by applying the PSO algorithm in Fig. 10, the optimal modulation variables $D_1$ and $D_2$ to achieve minimal current stress under chosen operating conditions $P$ and $V_2$ have been found. The configuration of PSO algorithm is given in Table IV. Fig. 14 (a), (b) and (c) graphically show the optimal $D_1$ and $D_2$ under the output voltage $V_2$ of 200V, 160V and 230V.

TABLE IV. CONFIGURATION OF PSO ALGORITHM

| Hyperparameter Name | Hyperparameter Value |
| --- | --- |
| Number of particles | 20 |
| Maximal number of iterations | 100 |
| Inertia weight $\omega$ | Linearly decrease from 0.9 to 0.4 |
| Learning coefficients $c_1$, $c_2$ | $c_1 = c_2 = 2.05$ |

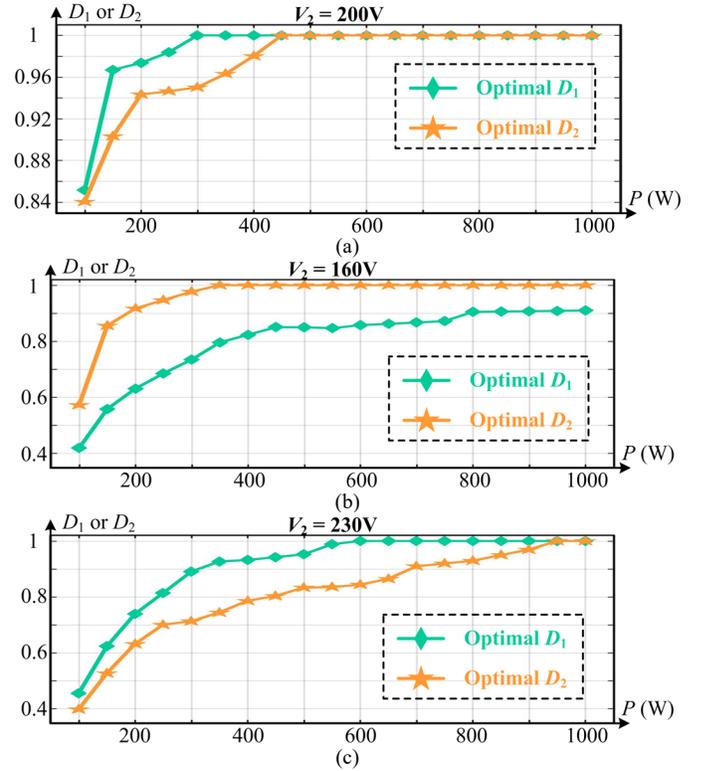

Fig. 14. Optimal $D_1$ and $D_2$ under different output voltage $V_2$ and different output power: (a) $V_2 = 200$ V; (b) $V_2 = 160$ V; (c) $V_2 = 230$ V.

### C. Stage III: Realization of TPS with FIS

TABLE V. CONFIGURATION OF FIS

| Inputs | $P$, $V_2$ of continuous values |
| --- | --- |
| Output | Optimal $D_1$, $D_2$ |
| Type of FIS | Type-1 Takagi-Sugeno [31] |
| Type of input membership functions | Gaussian |
| Number of input membership functions | 3 |
| Number of rules | 9 |

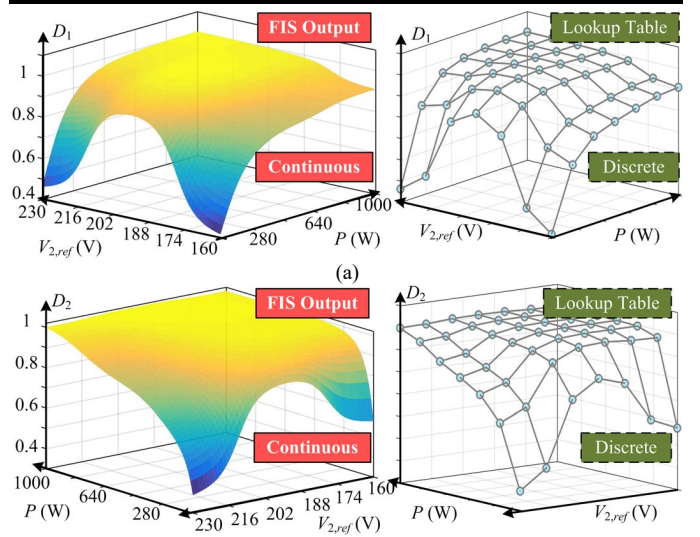

(b)

Fig. 15. Outputs of FIS with respect to the inputs $P$ and $V_2$: (a) $D_1$; (b) $D_2$.

In Stage III, the proposed FIS-based control diagram in Fig. 11 can realize optimal modulation variables $D_1$ and $D_2$ with minimal current stress under all possible continuous operating conditions $P$ and $V_2$. In this design example, type-1 Takagi-Sugeno FIS [31] is implemented, in which the input membership functions of $P$ and $V_2$ are gaussian and have three linguistic sets (low, medium and high). The configuration of FIS is summarized in Table V. The plots on the left side of Fig. 15 (a) and (b) show the outputs of FIS ($D_1$, $D_2$) with respect to the inputs $P$ and $V_2$ of continuous values, while those on the right are the exemplar discrete values saved in lookup table. As can be seen from Fig. 15, the proposed FIS-based TPS is superior to lookup table, achieving real-time modulation under all possibilities of $P$ and $V_2$.

### D. Computational Resources to Apply the Proposed AI-TPSM Approach in the Design Case

To provide insights of the computational resources required to apply AI-TPSM in the design case, the average CPU time of Stage I and II and the average turnaround time and storage size of Stage III are evaluated and shown in Table VI.

Compared to Stage II, running simulation and training NN in Stage I require more CPU time and resources. In Stage III, the turnaround time of deploying FIS online in the control platform Dspace 1202 is only 4.64 $\mu$s, indicating fast computation speed of FIS, and the storage size of FIS is only 3.4 $k$B, validating the low space complexity of FIS. As a comparison, if lookup table is applied online, the storage size can be large as several MB, which leads to unacceptable turnaround time [33].

TABLE VI. COMPUTATIONAL RESOURCES TO APPLY AI-TPSM

| | Platform | Performance |
|---|---|---|
| Stage I | Intel Xeon CPU E5-1630 @ 3.7 GHz, 16 GB RAM, Windows 10 | *Average CPU Time*: 3 days and 21 hours with four CPU cores |
| Stage II | | *Average CPU Time*: 1.57 hours |
| Stage III | Dspace 1202 | *Average Turnaround Time*: 4.64 $\mu$s *Storage Size*: 3.4 $k$B |

## V. EXPERIMENTAL VERIFICATION

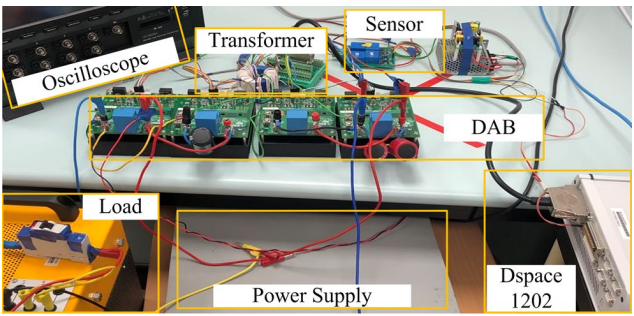

Fig. 16. Hardware platform in the experiments.

In this section, experiments have been conducted to validate the proposed AI-TPSM approach in the realization of optimal TPS modulation with minimal current stress in the design case. In the hardware experiments, design specifications are given in Table I, and the hardware platform is shown in Fig. 16.

The experiments in this section include the following parts: rated operation, operation under different $P$ and $V_2$, load and voltage step response, comparisons among SPS modulation, lookup-table-based TPS modulation and AI-TPSM, and comparisons between the experimental and theoretical results of the optimal modulation.

### A. Rated Operating Waveforms

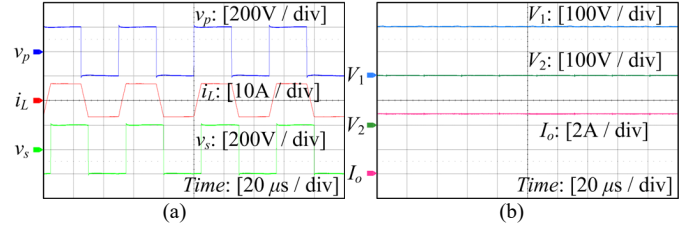

Fig. 17. Experimental waveforms under rated power of 1000W and rated output voltage of 200V: (a) $v_p$, $v_s$, and $i_L$; (b) $V_1$, $V_2$ and $I_o$.

Fig. 17 presents the rated operating waveforms of the DAB converter under the optimal TPS modulation with minimal current stress, which is designed by the proposed AI-TPSM approach. The notations and directions of the measured waveforms are given in Fig. 1. Under rated conditions, optimal $D_1$ and $D_2$ are both 1, so the waveforms $v_p$ and $v_s$ are purely square waves.

### B. Operating Waveforms under Different Output Power $P$ and Output Voltage $V_2$

In this part, experiments under output power $P$ of 100W, 400W and 900W and output voltage $V_2$ of 200V, 160V and 230V have been conducted.

Under $V_2$ of 200V, the waveforms $v_p$, $v_s$, and $i_L$ under 900W, 400W and 100W are given in Fig. 18, the working modes of which are mode 1, mode 1 and mode 5, respectively.

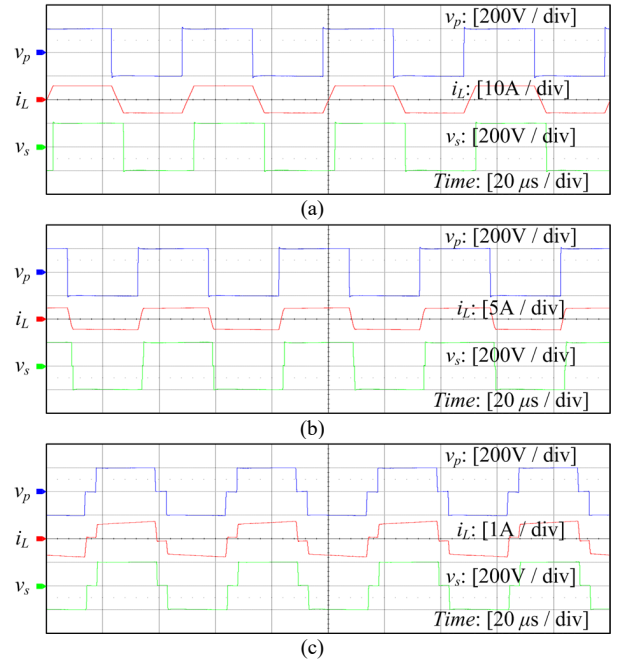

Fig. 18. Experimental waveforms under different output power when output voltage is 200V: (a) $P$ = 900W; (b) $P$ = 400W; (c) $P$ = 100W.

When $V_2$ is 160V, the waveforms $v_p$, $v_s$, and $i_L$ under 900W, 400W and 100W are shown in Fig. 19, the working modes of which are mode 1, mode 1 and mode 4, respectively.

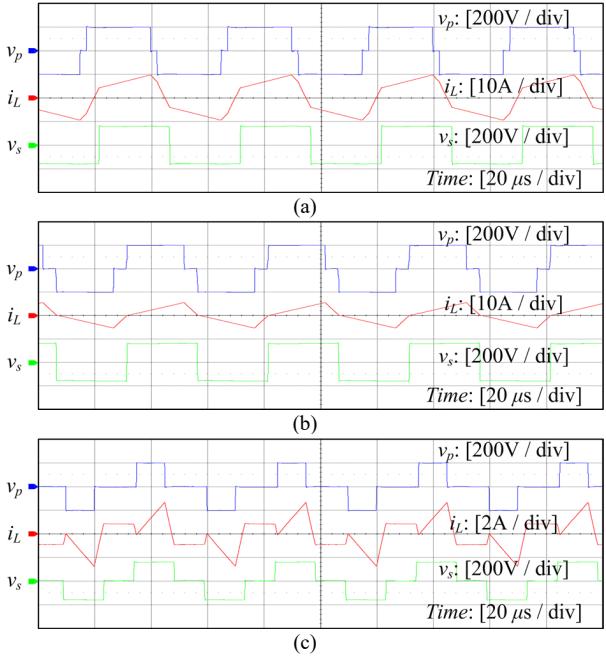

Fig. 19. Experimental waveforms under different output power when output voltage is 160V: (a) $P = 900$W; (b) $P = 400$W; (c) $P = 100$W.

Given $V_2$ of 230V, the waveforms $v_p$, $v_s$, and $i_L$ under 900W, 400W and 100W are given in Fig. 20, the working modes of which are mode 1, mode 5 and mode 5, respectively.

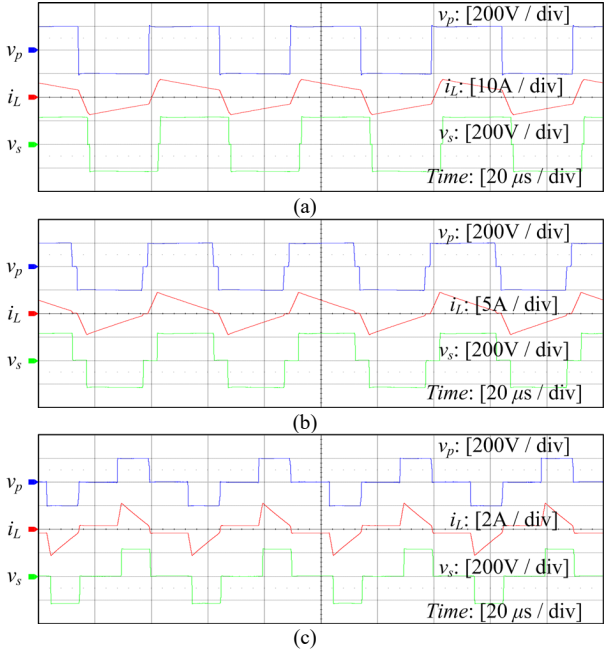

Fig. 20. Experimental waveforms under different output power when output voltage is 230V: (a) $P = 900$W; (b) $P = 400$W; (c) $P = 100$W.

### C. Transient Response under Power and Voltage Step

The experiments below aim at validating the real-time operation of the proposed FIS-based TPS modulation given the output power or voltage steps.

Firstly, when the load resistance is fixed at 40Ω, the output voltage $V_2$ steps from 200V (1000W) to 160V (640W) and from 160V (640W) to 200V (1000W). Fig. 21 shows the corresponding waveforms, where the top figures present the waveforms $V_1$, $V_2$ and $I_o$, and the bottom figures present the modulation waveforms $v_p$, $v_s$ and $i_L$.

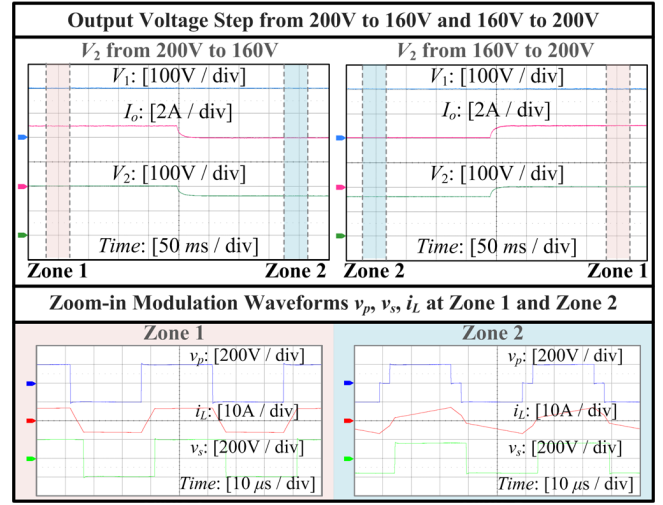

Fig. 21. Experimental waveforms given that $V_2$ steps from 200V to 160V and $V_2$ steps from 160V to 200V: $V_1$, $V_2$ and $I_o$ during voltage step (top); zoom-in view of $v_p$, $v_s$ and $i_L$ at Zone 1 and Zone 2 (bottom).

Secondly, when the load resistance is at 52.9Ω, the output voltage $V_2$ steps from 230V (1000W) to 200V (756W) and from 200V (756W) to 230V (1000W). Fig. 22 plots the waveforms.

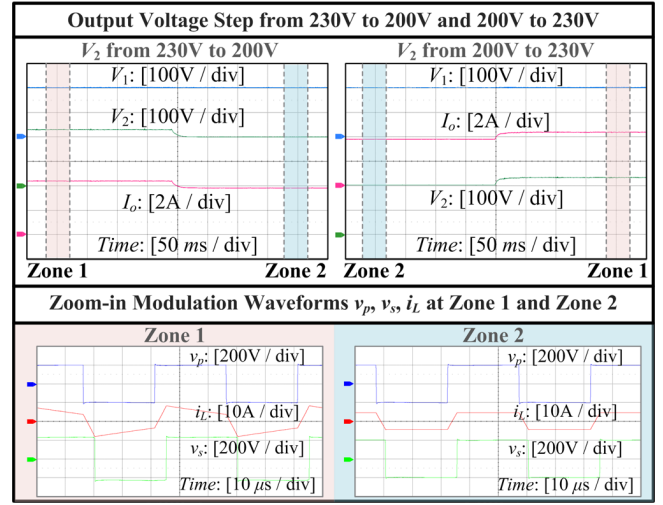

Fig. 22. Experimental waveforms given that $V_2$ steps from 230V to 200V and $V_2$ steps from 200V to 230V: $V_1$, $V_2$ and $I_o$ during voltage step (top); zoom-in view of $v_p$, $v_s$ and $i_L$ at Zone 1 and Zone 2 (bottom).

Additionally, Fig. 23, Fig. 24 and Fig. 25 present the transient responses of load steps under the output voltage of 200V, 160V and 230V, respectively.

From the waveforms shown in Fig. 21 and Fig. 22, when the operating conditions $V_2$ and $P$ vary, the proposed FIS-based TPS modulator can realize the real-time adjustments of $D_1$ and $D_2$ to achieve minimal current stress. Besides, when the output power $P$ varies, as shown in Fig. 23 to Fig. 25, the output voltage $V_2$ is capable of tracking the required reference value, and $D_0$, $D_1$, $D_2$ have been adjusted to their optimal values.

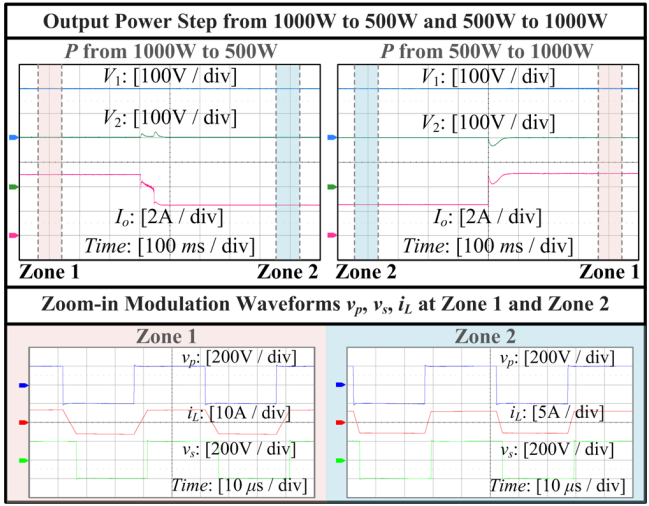

Fig. 23. Experimental waveforms given that $P$ steps from 1000W to 500W and from 500W to 1000W under $V_2$ of 200V: $V_1$, $V_2$ and $I_o$ during power step (top); zoom-in view of $v_p$, $v_s$ and $i_L$ at Zone 1 and Zone 2 (bottom).

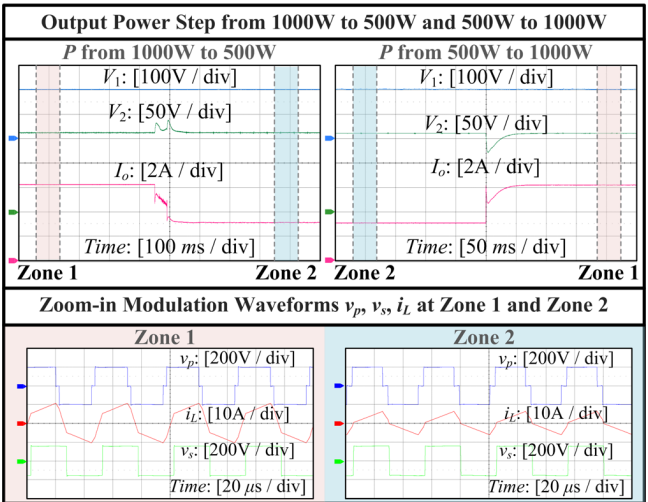

Fig. 24. Experimental waveforms given that $P$ steps from 1000W to 500W and from 500W to 1000W under $V_2$ of 160V: $V_1$, $V_2$ and $I_o$ during power step (top); zoom-in view of $v_p$, $v_s$ and $i_L$ at Zone 1 and Zone 2 (bottom).

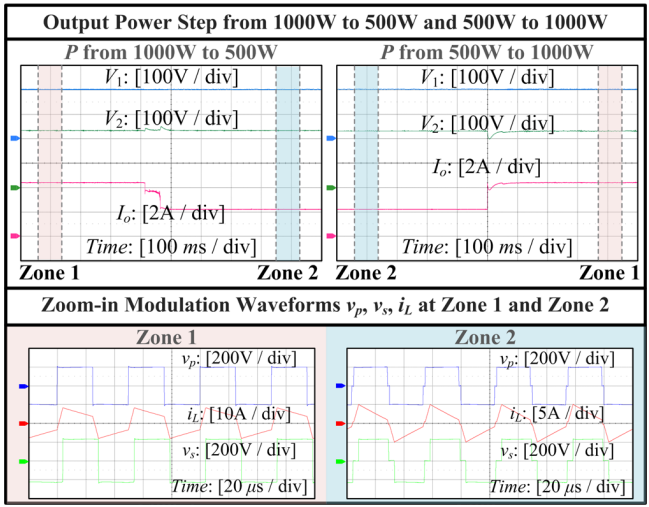

Fig. 25. Experimental waveforms given that $P$ steps from 1000W to 500W and from 500W to 1000W under $V_2$ of 230V: $V_1$, $V_2$ and $i_L$ during power step (top); zoom-in view of $v_p$, $v_s$ and $i_L$ at Zone 1 and Zone 2 (bottom).

In a word, with the conducted experiments, the proposed FIS-based TPS modulation is validated to work online under varying operating conditions.

### D. Current Stress and Efficiency Performance of the Optimal TPS Modulation via the Proposed AI-TPSM

To validate the satisfactory current stress and efficiency performance of the optimal TPS modulation via AI-TPSM, the conventional SPS modulation (SPSM) and the lookup-table-based TPS modulation (LUT-TPSM) are compared with. Fig. 26, Fig. 27 and Fig. 28 present the current stress $i_{pk}$ and efficiency $\eta$ from 100W to 1000W given that $V_2$ is 200V, 160V and 230V, respectively. In the experiments, efficiency is evaluated by the ratio of output power $P_o$ to input power $P_{in}$ utilizing the Teledyne LeCroy HDO8058A oscilloscope.

In the comparisons between SPSM and AI-TPSM, the optimal TPS modulation via the proposed AI-TPSM approach presents significantly lower $i_{pk}$ and higher $\eta$ at low power level. Apart from that, at medium power level, AI-TPSM also performs better in both $i_{pk}$ and $\eta$. At high power level, the conventional SPSM achieves almost the same performance as AI-TPSM. This relatively small advantage of AI-TPSM at high power level is because that the optimal $D_1$ and $D_2$ in these situations are close to 1, which renders the difference between the control signals of AI-TPSM and SPSM to be trivial.

In the comparisons between LUT-TPSM and AI-TPSM, the proposed AI-TPSM still achieves lower $i_{pk}$ and higher $\eta$ at low power levels. This superiority is benefited from the continuous modulation feature of AI-TPSM. Even though the lookup table in LUT-TPSM stores the optimal modulation variables, its results are still suboptimal due to its discrete modulation nature.

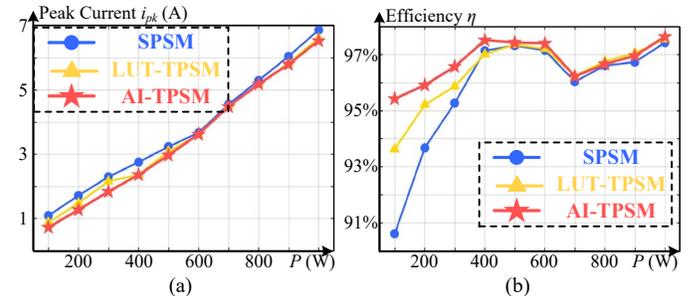

Fig. 26. Current stress $i_{pk}$ and efficiency $\eta$ performance of SPSM, LUT-TPSM and the optimal TPS modulation via AI-TPSM when output voltage $V_2$ is 200V: (a) current stress $i_{pk}$ performance; (b) efficiency $\eta$ performance.

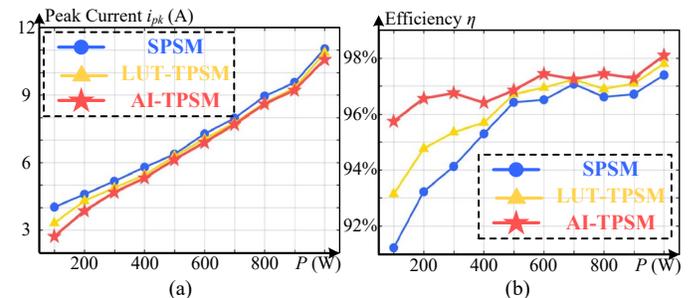

Fig. 27. Current stress $i_{pk}$ and efficiency $\eta$ performance of SPSM, LUT-TPSM and the optimal TPS modulation via AI-TPSM when output voltage $V_2$ is 160V: (a) current stress $i_{pk}$ performance; (b) efficiency $\eta$ performance.

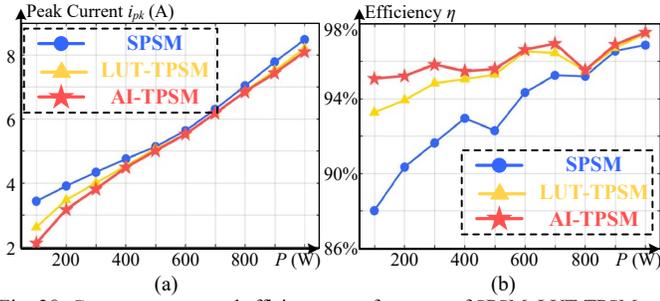

Fig. 28. Current stress $i_{pk}$ and efficiency $\eta$ performance of SPSM, LUT-TPSM and the optimal TPS modulation via AI-TPSM when output voltage $V_2$ is 230V: (a) current stress $i_{pk}$ performance; (b) efficiency $\eta$ performance.

In this part, in the comparisons among SPSM, LUT-TPSM and AI-TPSM, the superior current stress and efficiency performance of the proposed AI-TPSM approach is experimentally validated.

### E. Comparisons between the Experimental and Theoretical Results of the Optimal Modulation

To validate the high accuracy and optimality of the proposed AI-TPSM approach, the experimental results and the theoretically optimal results are compared.

As shown in Fig. 29, the average deviations between the experimental and theoretical results are 2.96%, 3.51% and 3.37% when the output voltage is 200 V, 160 V and 230 V, respectively. In the worst case, the largest deviation error is only 5.21%. Hence, the neglectable deviations from the experimental results to the theoretically optimal results validate the high accuracy of the proposed AI-TPSM, and they also prove that the experimental peak current results are indeed optimal.

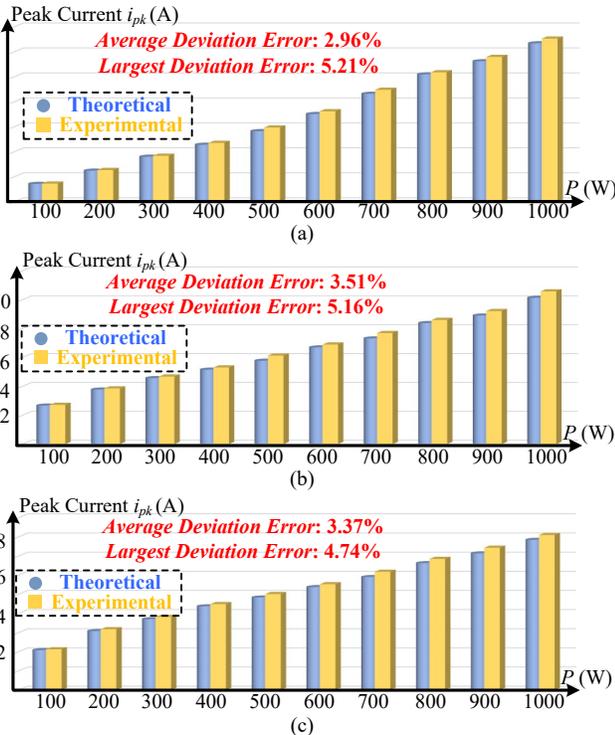

Fig. 29. Comparisons between the experimental results and the theoretically optimal results: (a) $V_2 = 200$V; (b) $V_2 = 160$V; (c) $V_2 = 230$V.

In a word, with all the hardware experimental results discussed above, the design case is comprehensively verified, and the proposed AI-TPSM approach for optimal TPS modulation with minimized current stress is validated.

## VI. CONCLUSION

This paper proposes an artificial-intelligence-based triple phase shift modulation optimization strategy (AI-TPSM) for the dual active bridge converter, which can realize minimized current stress in an automatic fashion. Generally, AI-TPSM can be divided into three stages with one AI tool in every stage: analysis process with neural network, optimization process with evolutionary algorithm and realization process with fuzzy inference system. Firstly, in analysis process, neural network is trained with simulations to learn the relationships between current stress and variables (operating conditions and modulation variables), which is to replace traditional complicated and inaccurate analytical process. Secondly, particle swarm optimization, which is an evolutionary algorithm, is adopted find the optimal modulation results which can minimize current stress. Lastly, fuzzy inference system is utilized to store the optimal modulation results under different operating conditions, which can provide continuous modulation. The proposed AI-TPSM enjoys high degree of automation which can relieve engineers' working burden and improve accuracy. Finally, the effectiveness of the AI-TPSM has been verified with a 1kW prototype of the DAB converter.

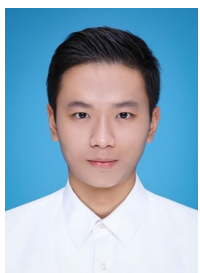
**Xinze Li** received his bachelor's degree in Electrical Engineering and its Automation from Shandong University, China, 2018. He has been awarded the Ph.D. degree in Electrical and Electronic Engineering from Nanyang Technological University, Singapore, 2023.

His research interests include dc-dc converters, modulation design, condition monitoring, digital twins for power electronics systems, design process automation, light and explainable AI for power electronics with physics-informed systems, application of AI in power electronics, and deep learning and machine learning algorithms.

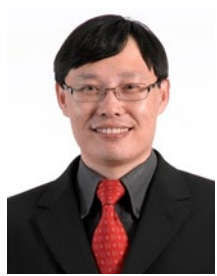
**Dr. Kezhi Mao** obtained his BEng, MEng and PhD from Jinan University, Northeastern University, and University of Sheffield in 1989, 1992 and 1998 respectively. He joined School of Electrical and Electronic Engineering, Nanyang Technological University, Singapore in 1998, where he is now a tenured Associate Professor.

Dr. Mao has over 20 years of research experience in artificial intelligence, machine learning, image processing, natural language processing, information fusion and cognitive science etc. He has published over 100 research papers in referred international journals and conferences. He has also edited 3 books published by Springer.

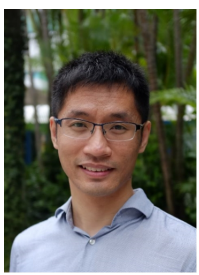
**Dr. Xin Zhang** received the Ph.D. degree in Automatic Control and Systems Engineering from the University of Sheffield, U.K., in 2016 and the Ph.D. degree in Electronic and Electrical Engineering from Nanjing University of Aeronautics & Astronautics, China, in 2014.

From February 2017 to December 2020, he was an Assistant Professor of power engineering with the School of Electrical and Electronic Engineering, Nanyang Technological University, Singapore. Currently, he is the professor at Zhejiang University. He is generally interested in power electronics, power systems, and advanced control theory, together with their applications in various sectors.

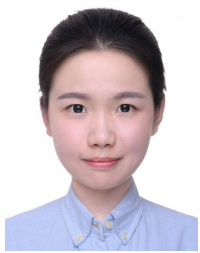
**Fanfan Lin** was born in Fujian, China in 1996. She received her bachelor degree in electrical engineering from Harbin Institute of Technology in China in 2018. From 2018, she has been awarded the Joint Ph. D. degree in Nanyang Technological University, Singapore and Technical University of Denmark, Denmark. Her research interest includes large language models for the design of large-scale power electronics systems, multi-modal AI for the maintenance of power converters, and the application of AI in power electronics.

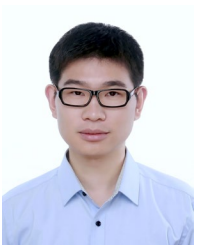
**Dr. Changjiang Sun** (S'13–M'19) received the Ph.D. degree from Shanghai Jiao Tong University, China, in 2019.

He is currently a Research Fellow with Nanyang Technological University, Singapore. His current research interests include topology and control of dc-dc converters for application in renewable energy collection and transmission grids.